\documentstyle[pra,aps,floats]{revtex}
\renewcommand{\cite}[1]{$^{(#1)}$}
\newcommand{\be}{\begin{equation}}
\newcommand{\ee}{\end{equation}}  
\newcommand{\ba}{\begin{eqnarray}}
\newcommand{\ea}{\end{eqnarray}}  
\newcommand{\de}{\partial}
\input epsf

\begin{document}
\twocolumn[\hsize\textwidth\columnwidth\hsize\csname
@twocolumnfalse\endcsname

\title{Transient Chaos and Critical States in Generalized Baker Maps
}
\author{Z. Kaufmann$^{1}$ and P. Sz\'epfalusy$^{1,2}$}
\address{
$^1$Department of Physics of Complex Systems,
E\"otv\"os University,
P. O. Box 32, H-1518 Budapest, Hungary\\
$^2$Research Institute for Solid State Physics and Optics,
P. O. Box 49, H-1525 Budapest, Hungary}
\maketitle

\begin{abstract}%
Generalized multibaker maps are introduced to model dissipative systems which
are spatially extended only in certain directions and escape of particles is
allowed in other ones.
Effects of nonlinearity are investigated by varying a control parameter.
Emphasis is put on the appearance of the critical state
representing the borderline of transient chaos,
where anomalous behavior sets in.
The investigations extend to the conditionally invariant and the related
natural measures and to transient diffusion in normal and critical states as
well.
Permanent chaos is also considered as a special case.

\end{abstract}
\vskip2pc] 

\narrowtext

\section{Introduction}

Transient chaos has become of equal importance as the permanent one in the
last decade.
Its field of applications has included chaotic scattering, chaotic advection
and transport phenomena\cite{1-9}.
The conditionally invariant measure\cite{10,6} 
and the related natural one\cite{11,6} 
are the counterparts of the invariant measure existing in systems exhibiting
permanent chaos.
Statistical properties of the trajectories during their chaotic development
can be expressed with the help of these measures.
In particular the properties of diffusion in systems whose trajectories can
escape from the phase space considered is governed by the natural measure.
The main purpose of the present paper is to further explore the features of
measures attached to transient chaos in dissipative systems in particular to
follow their change when a control parameter is varied up to the borderline
where the transient chaos shows critical behavior.
The motivation of the present paper stems from the features of deterministic
diffusion in transiently chaotic systems\cite{7}. 
Its peculiar properties can be traced back to those of the conditionally
invariant measures and the related natural ones.
In this problem two main areas of chaotic phenomena are present simultaneously,
namely transient chaos and deterministic diffusion, which has played a central
role in understanding macroscopic transport properties from the point of view
of deterministic microscopic motion
since the pioneering work by Gaspard and Nicolis\cite{1}. 

Deterministic diffusion has mainly been studied within the framework of
two models, namely considering an infinite chain of one-dimensional
maps\cite{12-16,1,4,7}
or with the help of multibaker maps\cite{2,3,5}. 
In such systems due to translational invariance the statistical properties of
chaos can be investigated within the unit cell by introducing
the reduced map.
In case of the dyadic multibaker map\cite{3} 
the reduced map agrees with the original Baker map.
Needless to say that the reduced map can be interesting in itself as a
two-dimensional map exhibiting various chaotic behavior independently from its
relationship to an infinite chain of similar maps.

In our previous paper we extended models of the first type
to allow escape at each step, which results in a reduced map with a window,
where the trajectory can leave the system\cite{7}. 
To study transient diffusion further the present paper generalizes
the dyadic multibaker map in two respects.
Namely, we introduce again a window in the reduced map and furthermore treat
nonlinear two-dimensional maps instead of the piecewise linear baker map.
The latter extension
makes it possible to follow the parameter dependence of characteristics
from the baker map up to the borderline situation,
where hyperbolicity is violated and
the limiting value of the escape rate agrees with the positive
Lyapunov exponent of the saddle point at the origin.
Since we want to study dissipative systems, the maps treated are invertible
but not invariant under time-reversal,
i.e.\ taking the opposite time direction we do not get the same map or its
conjugate.
(See Ref.\ 3 
for general definition of time-reversal invariance and for the proof that
the original multibaker map fulfills the requirements.
A time-reversible generalization of the Baker map is given in Ref.\ 8.)

The reduced maps considered have conditionally invariant measures which are
smooth along the unstable manifold.
It is pointed out that while the conditionally invariant
measures\cite{10,17,18} 
are different 
for forward and backward iterations, the related natural measures
are the same, which
ensures that the diffusion coefficient has the same value in both directions.
Approaching the borderline situation\cite{19,20} 
the natural measure
(but not the conditionally invariant one) degenerates to a Dirac delta function
located at the origin, which has the consequence that the diffusion coefficient
tends to zero in this limit.
At the borderline situation, however, further conditionally invariant 
measures smooth in the unstable direction appear,
which have, of coarse, different basins of attraction.
The type of the latter ones may generate a non-degenerate natural measure
and a finite diffusion coefficient in one of the directions of iterations.

In the permanently chaotic case anomalous diffusion\cite{15,16,21,22}
can occur in the strongly
intermittent state (when the reduced map with the considered measure has
Kolmogorov-Sinai entropy $K=0$).
Following the borderline situations, where the transient chaos is critical
the limit when the escape goes to zero can lead to weakly ($K>0$) or strongly
intermittent dynamics.
In the former case there remains no anomalous behavior in the diffusion.

The paper is organized as follows.
In Section II the model to be investigated is introduced and its properties
are discussed. 
The invariant set 
(chaotic saddle) of the reduced maps is specified,
and the Frobenius-Perron operator, the conditionally invariant measure,
and the natural measure are defined and studied in Section III\@.
The special case of the critical (borderline) situation is treated in
Section IV\@.
Properties of transient diffusion are considered in Section V\@.
Section VI is devoted to a discussion.

\section{Generalized Multibaker Maps}

We introduce the two-dimensional maps
$(x',y',S')=F(x,y,S)$ with the specification
\begin{equation}
F(x,y,S)=\left\{\begin{array}{lll}
(f(x),J^F y/|f'(x)|,S-1)   &\mbox{ if } &x\in {\rm I}_0\;,\\
(f(x),1-J^F y/|f'(x)|,S+1) &\mbox{ if } &x\in {\rm I}_1\;,\end{array}\right.
\label{mb}
\end{equation}
where {$S$ is an integer variable,}
$f(x)$ is a smooth function mapping $[0,1]$ twice onto 
itself,
${\rm I}_0=f_l^{-1}[0,1]$, ${\rm I}_1=f_u^{-1}[0,1]$ and
$f_l^{-1}$, $f_u^{-1}$ denote the lower and upper branches of the
inverse of $f(x)$, respectively.
We shall assume that $f(x)$ is increasing in ${\rm I}_0$ and
decreasing in ${\rm I}_1$.
It may have an escape window between ${\rm I}_0$ and ${\rm I}_1$,
for which it is not defined.

The map (\ref{mb}) has a constant Jacobian $J^F>0$.
It acts on an infinite chain of unit squares.
$S$ labels the square which is visited by the particle and $(x,y)$
determines the point inside the square.
During the mapping the particle jumps to the square to the right if
$x\in {\rm I}_1$ or to the left if $x\in {\rm I}_0$.
If there is an escape window in $f$ then there is a third possibility,
the particle leaves the system if
$x\in [0,1]\setminus{\rm I}_0\setminus{\rm I}_1$.

Translational invariance of the system makes it possible to use the so called
reduced map for a partial description of the system\cite{12-14}.
It can be obtained by considering the cells to be
identical, and thereby, following only the motion inside one cell.
The reduced map $F(x,y)$ is the same as $F(x,y,S)$ disregarding the 
variable $S$.

The reduced map of (\ref{mb}) maps two rectangles of the unit square
${\rm U}=[0,1]\times [0,1]$ in the $(x,y)$ plane to two regions
%
(see Fig.\ 1),
which can in general overlap.
To exclude this overlap, which causes problem concerning invertibility, we
choose the 1D map $f(x)$ such, that the constant density $P(x)\equiv 1$ 
be conditionally invariant under its action.
$P(x)$ satisfies the one-dimensional
Frobenius-Perron equation
\begin{equation}
e^{-\kappa}P(x)=
\frac{P(f_l^{-1}(x))}{|f'(f_l^{-1}(x))|}
+\frac{P(f_u^{-1}(x))}{|f'(f_u^{-1}(x))|}\;,\label{fp1}
\end{equation}
where $\kappa$ is the escape rate.
Requiring that $P(x)\equiv 1$ be a solution of Eq.\ (\ref{fp1}) by integration
and taking into account the properties of $f(x)$ listed below Eq.\ (\ref{mb})
one obtains
\be
f^{-1}_u(x)-f^{-1}_l(x)=1-e^{-\kappa}x\;.\label{fidiff}
\ee
In the following it will be understood that $f(x)$ satisfies this requirement.
Note that such a 
condition does not restrict generality in the choice of the 1D map, since
any map $f(x)$, as specified below (\ref{mb}),
with smooth conditionally invariant measure
can be transformed to a map with constant conditionally invariant
density\cite{23,19}. 
The necessary transformation is a conjugation $\mu(f(\mu^{-1}(x)))$,
where $\mu(x)$ is the conditionally invariant measure of $[0,x]$ for the map 
$f(x)$.

\begin{figure}
\begin{center}\epsfbox{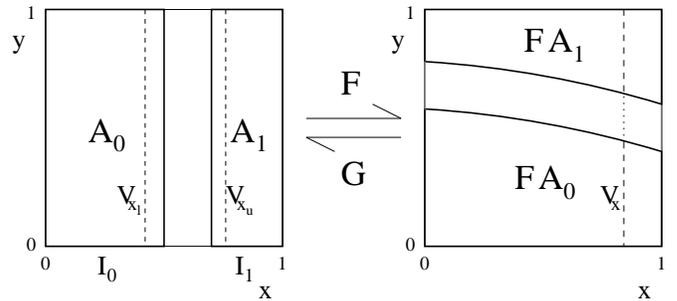}\end{center}
\caption{Schematic drawing of the action of the reduced maps $F$ and $G$.}
\end{figure}

The general form of such a map can be given with the formulas for its
inverse\cite{23,19} 
as follows
\begin{eqnarray}
f_l^{-1}(x)&=&\frac{x}{2R}+v\left(\frac{x}{2R}\right)\;,\label{fl1}\\
f_u^{-1}(x)&=&1-\frac{x}{2R}+v\left(\frac{x}{2R}\right)\;.\label{fl2}
\end{eqnarray}
Here $v(x)$ is a smooth function with properties $v(0)=0$ and $|v'(x)|\le 1$, 
where equality is allowed only in isolated points.
Substituting (\ref{fl1},\ref{fl2}) into (\ref{fp1}) one obtains
$R=e^\kappa$.
Note that $v=0$ leads to the tent map.
With the choice $v(x)=d\cdot x(1-x)$, $d\in[-1,1]$
Eqs.\ (\ref{fl1},\ref{fl2}) specify a one-dimensional map which is conjugated
to the one used in Ref.\ 7. 

To obtain condition for invertibility of $F$ let us determine the preimage
of a vertical line segment ${\rm V}_{x}=\{(x,y)\,|\;y\in[0,1]\}$.
It is reached from two vertical segments
${\rm V}_{x_l}$ and ${\rm V}_{x_u}$.
%
See Fig.\ 1, where the segment of $V_x$ which does not have a preimage
is shown by dotted line.
The corresponding $x$ values are $x_l=f_l^{-1}(x)$ and
$x_u=f_u^{-1}(x)$, respectively.
The first segment, ${\rm V}_{x_l}$ is mapped to the subinterval
$y\in[0,J^F/f'(f_l^{-1}(x))]$,
and ${\rm V}_{x_u}$ to $y\in[1-J^F/|f'(f_u^{-1}(x))|,1]$ on ${\rm V}_{x}$.
{}From (\ref{fp1}) follows, that these subintervals do not overlap, if
\begin{equation}
J^F e^{-\kappa}\le 1\;,\label{co}
\end{equation}
where $\kappa$ is defined by (\ref{fp1}) with $P(x)\equiv 1$.
In the following this condition shall be assumed to be satisfied.

As $x$ runs through the interval $[0,1]$ 
the subintervals mentioned above fill in two separate regions,
namely the images $F{\rm A}_0$, $F{\rm A}_1$
of ${\rm A}_0={\rm I}_0\times [0,1]$ and ${\rm A}_1={\rm I}_1\times [0,1]$%
%
(see Fig.\ 1).
A special feature of the map with $v(x)=d x(1-x)$ is, that the boundaries of 
$F{\rm A}_0$ and $F{\rm A}_1$ are straight lines.
However, further iterates have curved boundaries for $d\neq 0$ even in this 
case.
On the other hand the length of the gap between the two intervals
$[0,J^F/f'(f_l^{-1}(x))]$ and $[1-J^F/|f'(f_u^{-1}(x))|,1]$
is $1-J^Fe^{-\kappa}$, which is independent of $x$, yielding the occurrence of
an empty stripe in the unit square with constant vertical thickness
if $J^F e^{-\kappa}<1$.
Even in the limiting case when the thickness becomes zero (i.e.\ equality 
holds in (\ref{co})) the
common points of $F{\rm A}_0$ and $F{\rm A}_1$
form a set of measure zero,
such as in the case of the original baker map, and therefore they can be 
disregarded.
Thereby the reduced map, and also (\ref{mb}) are invertible
for typical points.
The inverse map reads
\begin{equation}
G(x,y,S)=\left\{\begin{array}{l}
(f_l^{-1}(x),J^G f'(f_l^{-1}(x)) y    ,S+1)\\
\hspace{2cm}\mbox{ if } y\le\frac{1}{J^G f'(f_l^{-1}(x))}\;, \\
(f_u^{-1}(x),J^G|f'(f_u^{-1}(x))|(1-y),S-1)\\
\hspace{2cm}\mbox{ if } y\ge 1-\frac{1}{J^G|f'(f_u^{-1}(x))|}\;,
\end{array}\right.\label{mbi}
\end{equation}
where $J^G=(J^F)^{-1}$ is the Jacobian of the iteration $G$.
Similarly to $F$, $G$ maps points of a square to the neighboring cells.
Namely, the point jumps to the right if $(x,y)\in F {\rm A}_0$
and to the left if $(x,y)\in F {\rm A}_1$.

For the iteration $F$ the range $J^F\in(0,1)$ is the physically 
interesting one.
However, if we consider the iteration $G$ to be the physical 
direction, then $J^G\in(0,1)$ is the relevant one.
While in case of iteration $F$ one of the variables, namely $x$, 
transforms independently from the other one, it is no more true in case of the
iteration $G$.
This means that the two-dimensional features of the map are more pronounced in
the latter case.
We consider $F$ and $G$ on equal footing.

The maps (\ref{mb}) and (\ref{mbi})
have the advantage that their reduced maps,
which are generalized baker maps,
have complete grammar if one chooses the symbolic dynamics
generated by the partition $({\rm A}_0,{\rm A}_1)$.
We note, that similar maps have been studied in Ref.\ 24 
in other context.

Eq.\ (\ref{mb}) contains as special cases piecewise linear maps by choosing
$f(x)=ax$ if $x<1/a$ and $f(x)=a(1-x)$ if $x\ge 1-1/a$,
or, alternatively, $f(x)=ax$ if $x<1/a$ and $f(x)=1-a(1-x)$ if $x\ge 1-1/a$.
The latter one with the choice $a=2$ gives back the original dyadic multibaker 
map\cite{3}. 

\section{The Invariant Set and Invariant Measures}

To discuss invariant measures one has to study evolution equation of a 
probability distribution under the effect of the reduced map.
The Frobenius-Perron operator transforming the density of a smooth measure
in two dimensions can be written for the map $F(x,y)$ as
\begin{equation}
L P(x,y)\equiv\left\{\begin{array}{ll}
P(F^{-1}(x,y))/J^F &\mbox{ if } F^{-1}(x,y)\in {\rm U},\\
0 &\mbox{ otherwise,}
\end{array}\right.\label{fp2}
\end{equation}
with ${\rm U}=[0,1]\times[0,1]$ and similarly for the map $G$.

Consider first the map $F$ with escape.
The conditionally invariant measure $\mu^F$ is defined as the limit of the
measures
with densities $\exp(\kappa^F n)L^n P_0(x,y)$ when $n\rightarrow\infty$,
where $\kappa^F$ is the escape rate related to $F$.
It is easy to see, that in the unstable direction the conditionally invariant
measure is smooth.
Since iteration of $x$ by the reduced map is independent of $y$, the projection
of the conditionally invariant measure to the $x$ axis is equal to the 
conditionally invariant measure of the 1D map, which has constant density.
It also follows that the corresponding escape rate $\kappa^F$ equals to the
escape rate $\kappa$ of the 1D map $f$.
On the other hand, from the appearance of the empty stripe it is obvious, 
that the conditionally invariant measure has a fractal basis in the $y$ 
direction, if $J^F\exp(-\kappa^F)<1$.
%
Consequently it does not have a smooth density and can be best demonstrated
by derivating it only in $x$ direction, i.e.\ by displaying
$\mu^F_x(x,y)=\frac{\de}{\de x}\mu^F([0,x]\times[0,y])$.
When there is a density it is connected to $\mu^F_x$ as
$P^F(x,y)=\frac{\de}{\de y}\mu^F_x(x,y)$.
However, in the present case $\mu^F_x$ is
a devil's staircase in three dimensions, as seen in Fig.\ 2a
in a finite resolution,
and $P^F$ is a bounded function only in numerical approximation.
In area preserving case there should be an escape to 
fulfill this inequality, while in the dissipative case ($J^F<1$) escape is not
necessary.
In case of area expanding maps
(when $G$ is physically relevant)
$\kappa^F>\log J^F$ should stand.
In the case $J^F\exp(-\kappa^F)=1$ starting from a uniform density
$P_0(x,y)=1$ the full square is filled uniformly after one mapping,
because of the constant Jacobi determinant.
Then there are still two possibilities.
If $\kappa^F>0$ the resulting distribution gives back $P_0$ after
normalization, so the conditionally invariant measure is the Lebesgue measure.
If $\kappa^F=0$ then $J^F=1$ and the situation is the same but normalization
is not necessary.

\begin{figure}
\noindent\hspace*{-1cm}\epsfbox{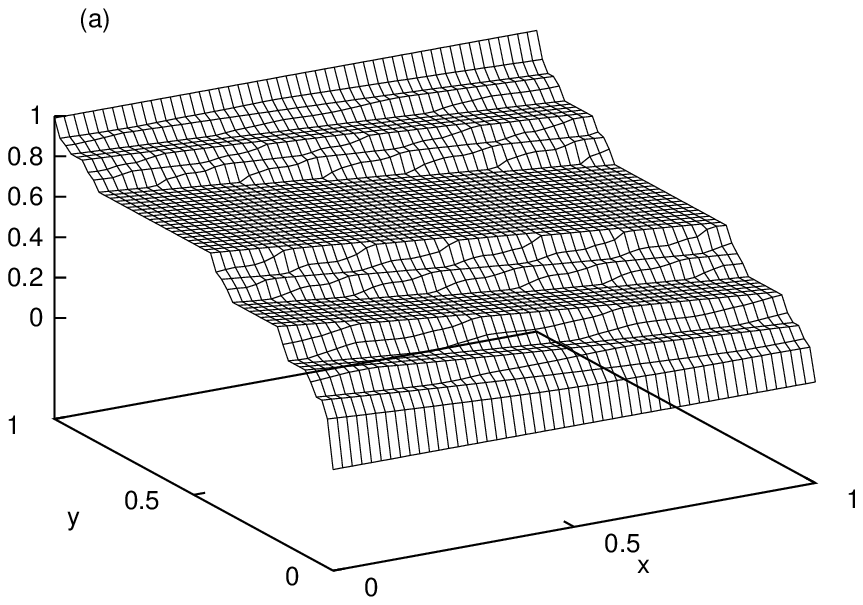}

\noindent\hspace*{-1cm}\epsfbox{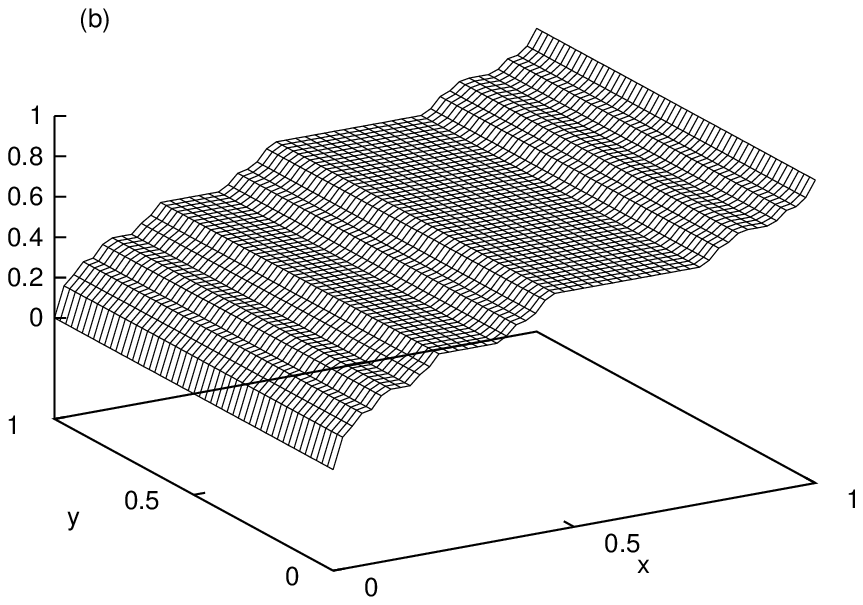}
\caption{(a) The derivative
$\mu^F_x(x,y)=\frac{\de}{\de x}\mu^F([0,x]\times[0,y])$ of the
conditionally invariant measure $\mu^F$ related to $F$
which is specified by Eqs.\ (4), (5) and
with choice $v(x)=d\cdot x(1-x)$, $R=1.2$, $J^F=0.7$, $d=0.5$.
(b) The derivative
$\mu^G_y(x,y)=\frac{\de}{\de y}\mu^G([0,x]\times[0,y])$ of the
conditionally invariant measure $\mu^G$ related to $G$
which is specified by Eqs.\ (4), (5) and
$v(x)=d\cdot x(1-x)$, $R=1.5$, $J^G=0.8$, $d=0.5$.}
\end{figure}

The natural measure is defined by the distribution of points of
trajectories that have been started $n$ iterations before 
according to an initial distribution with density $P_0(x,y)$
and stay in the system for at least another $n$ iterations,
also taking the limit $n\rightarrow\infty$.
The first $n$ steps map $P_0(x,y)$ to a smooth distribution in
$F^n{\rm U}\cap {\rm U}$, thereby approaching the conditionally invariant 
measure when normalized.
The restriction for the further $n$ steps selects the points in 
$F^{-n}{\rm U}$.
Since points that have left ${\rm U}$ are assumed not to return to it, the 
common part
${\rm U}_n=F^n{\rm U}\cap F^{-n}{\rm U}$ lies inside ${\rm U}$,
and gives the invariant set in the limit.
The above restrictions mean that the natural measure is obtained as the
limit of the iterated and normalized
distribution in ${\rm U}_n=F^n{\rm U}\cap F^{-n}{\rm U}$.
In the rest of this section we assume that the initial distribution is
according to the Lebesgue measure.
Because the Jacobian is constant
the result of $n$ steps is the Lebesgue 
measure restricted to ${\rm U}_n$ (apart from normalization), and the natural 
measure is its limit for $n\rightarrow\infty$. 
The invariant set and the natural measure are fractal in both directions if 
$J^F\exp(-\kappa^F)<1$ and $\kappa^F>0$, 
since the first condition ensures presence of horizontal stripes in 
$F^n{\rm U}$ and the latter creates vertical strips in $F^{-n}{\rm U}$.
An exception is the critical situation, which will be discussed
in the next section.

Let us study now the iteration $G$ with escape.
Starting from $P_0(x,y)=1$, after $n$ iterations we obtain
\begin{equation}
P_n(x,y)=\left\{\begin{array}{ll}
(J^G)^{-n} &\mbox{ if }
f^n(x)\in[0,1]\;, \\
0 &\mbox{ otherwise}\;.
\end{array}\right.
\end{equation}
Together with normalization by factor $\exp(\kappa^G n)$ this yields the
density of the conditionally invariant measure $\mu^G$
in the limit $n\rightarrow\infty$, where $\kappa^G$ is the escape rate.
The $x$-dependence of $P_n(x,y)$ is,
apart from normalization, equivalent to the restriction
of the conditionally invariant density of the 1D map $f(x)$ to $f^{-n}[0,1]$.
So in the $x$ direction 
$P_n(x,y)$ gives in the limit $n\rightarrow\infty$ after normalization the 
natural measure of the 1D map $f(x)$.
Therefore the conditionally invariant measure of $G$ is 
equal to the natural measure of the 1D map $f(x)$ in $x$ direction and uniform 
in $y$ direction.
%
Fig.\ 2b illustrates this measure.
For reasons similar to case of map $F$ the derivative 
$\mu^G_y(x,y)=\frac{\de}{\de y}\mu^G([0,x]\times[0,y])$ is displayed.
When there is a density it is connected to $\mu^G_y$ as
$P^G(x,y)=\frac{\de}{\de x}\mu^G_y(x,y)$.
However, in the present case $\mu^G_y$ is a devil's staircase
and there is no bounded density.
Since the normalization factor in the natural measure of $f(x)$ is
$\exp(\kappa^F n)$ the normalization factor for $P_n$ is
$\exp(\kappa^F n)(J^G)^n$ and the corresponding escape rate is
$\kappa^G=\kappa^F-\log J^F$.
This connection can be written in a symmetrized form as
\begin{equation}
\kappa^F-\frac 12 \log J^F = \kappa^G-\frac 12 \log J^G\;.\label{kfg}
\end{equation}
Note that $\kappa^G$ is zero if the equality sign applies in Eq.\ (\ref{co}).

If we apply the above construction of the natural measure to the map $G$ we 
replace $n$ by $-n$ in ${\rm U}_n$ and obtain the same set.
Due to the constant Jacobian,
starting with the Lebesgue measure we obtain the uniform distribution in
${\rm U}_n$ for the approximation of the natural measure, as in case of map $F$.
Therefore their limits, the natural measures of $F$ and $G$ are identical.

It is interesting to study the special cases when
one of the escape rates, $\kappa^F$ or $\kappa^G$ is zero.
(Note, if both are zero the map is not dissipative, which case we do not want
to discuss.)
If $\kappa^F$ ($\kappa^G$) is zero
the map $F$ ($G$) exhibits permanent chaos.
Then the conditionally invariant measure $\mu^F$ ($\mu^G$)
similarly to the natural measure becomes invariant under the mapping $F$ ($G$).
This is a fractal measure (not smooth in the $y$ ($x$) direction)
since $\kappa^G$ ($\kappa^F$) is still positive.
On the other hand the iteration $G$ ($F$) maps a constant density
to a constant one in the square $\rm U$,
therefore the conditionally invariant measure $\mu^G$ ($\mu^F$)
is the Lebesgue measure in $\rm U$.

\section{Critical State}

We call a state critical if the invariant natural measure concentrates on a
non-fractal subset of the repeller
while the conditionally invariant measure is smooth along the unstable
direction when one starts with the Lebesgue measure.
In the examined 1D maps and in the present 2D ones this non-fractal subset
is a fixed point.
In 1D maps the borderline situation showing critical transient chaos
is achieved when the slope of the map at a fixed point 
equals\cite{19,20} 
$e^\kappa$.
In the representation, used here,
when the Lebesgue measure is a conditionally invariant one
this means that the slope of the map is infinity in the preimage of the 
fixed point, as can be seen from (\ref{fidiff}).
Towards this limit while the conditionally invariant measure remains smooth
the natural measure degenerates to a Dirac delta function at the fixed 
point\cite{19,20}. 
It occurs for maps (\ref{fl1},\ref{fl2}) for $v'(0)=1$.
In this case the inverse map (\ref{fl2}) behaves near $x=0$ as
\be
f_u^{-1}(x)-1\propto x^\omega\;,\label{odef}
\ee
where $\omega>1$.

In the 2D map (\ref{mb}) the condition for criticality,
as will be shown below, is that the positive 
Lyapunov exponent of the saddle point in the origin agrees with $\kappa^F$.
In terms of $v(x)$ for the map (\ref{mb}) with (\ref{fl1},\ref{fl2}) the 
condition again reads as $v'(0)=1$.
We can see that at criticality two
conditionally invariant measures $\mu^F_A, \mu^F_B$ have to be considered
possessing necessarily different escape rates\cite{7,25}. 
In case $v'(0)=1$ the side of $F {\rm A}_1$ at $x=0$
(which is the image of the $x=1, y\in[0,1]$ line segment)
shrinks to a point, as seen from (\ref{mb}) and (\ref{fl2}).
Since the Jacobian is constant, necessarily the modulus of the local
Lyapunov exponent becomes infinity in both directions.
The behavior of the measure can be characterized by the measure of
the rectangles ${\rm Box}(0,x)=[0,x]\times [0,1]$ and
${\rm Box}(x,1)=[x,1]\times [0,1]$.
Integrating (\ref{fp2}) over ${\rm Box}(0,x)$ one obtains
\be
L\mu({\rm Box}(0,x))=\mu({\rm Box}(0,f_0^{-1}(x)))
+\mu({\rm Box}(f_1^{-1}(x),1))\;.
\label{boxfp}
\ee
Therefore, starting with the Lebesgue measure,
when $\mu({\rm Box}(0,x))=x$,
it follows from (\ref{fidiff}) and (\ref{boxfp}) that
$L\mu({\rm Box}(0,x))=e^{-\kappa}x$.
Similarly, the property $\mu({\rm Box}(0,x))\propto x$ is kept in further
iterations, thereby the normalized conditionally invariant measure also
possesses the property $\mu^F_A({\rm Box}(0,x))=x$
with escape rate $\kappa^F=\kappa$.
%
Consequently
$\mu^F_{Ax}(x,y\!=\!1)=\frac{\de}{\de x}\mu^F_A({\rm Box}(0,x))$
is unity for every $x$, as seen in Fig.\ 3a.
On the other hand, inserting infinitesimally small $x$ into (\ref{boxfp}) one
obtains $L\mu^F_A({\rm Box}(0,x))=f^{-1}_l(x)x$,
therefore $\kappa^F_A=\log f'(0)$.
Starting now with a smooth distribution with scaling
$\mu({\rm Box}(0,x))\propto x^\omega$ for small $x$ due to (\ref{odef}) both 
terms
in (\ref{boxfp}) scale as $x^\omega$, leading to a conditionally invariant
measure for which $\mu^F_B({\rm Box}(0,x))\propto x^\omega$
%
and $\mu^F_{Bx}(x,y)=\frac{\de}{\de x}\mu^F_B([0,x]\times[0,y])
\propto x^{\omega-1}$ for small $x$.
Such a measure with $\omega=2$ is shown in Fig.\ 3b.

Consequently any initial distribution with density vanishing fast enough
at $x=0$ (and not diverging too fast at $x=1$) will approach $\mu^F_B$
when iterated keeping its norm constant.
It can be shown, similarly as for the case of 1D maps\cite{25}, 
that the condition is that
the measure $\mu^F_B({\rm Box}(0,x))$ decreases at least as fast as
$x^{\sigma_c}$ when $x\rightarrow 0$
where $\sigma_c={\kappa_B^F}/{\kappa_A^F}$
and $\mu^F_B({\rm Box}(x,1))$ decreases at least as $(1-x)^{\sigma_c/\omega}$
when $x\rightarrow 1$.

It is more surprising, that the criticality has serious consequences
also for the characteristics of the iteration $G$.
This appears already in the form of the conditionally invariant measure
$\mu^G$ of the map $G$ obtained starting with the Lebesgue
measure, when approaching criticality.
In this limit its density becomes a Dirac delta function in $x$ and constant
in $y$ direction,
as it follows from the considerations of the previous section
%
(see Fig.\ 3c).
Though the natural measure corresponding to $\mu^F_B$
is also invariant under action of $G$,
there is no counterpart of $\mu^F_B$ for the map $G$.
Instead, as it can be easily checked, at the critical state there is an
infinity of conditionally invariant measures $\mu^G_\eta$
smooth in the $y$ direction.
They are concentrated on the $x=0$ line segment, similarly to $\mu^G$.
Namely, they have the density $y^\eta\delta(x)$ with $\eta>-1$,
and the escape rates $\kappa^G_\eta=(\eta+1)\kappa^G$ belongs to them.
Among them the case $\eta=0$ represents the conditionally invariant measure
obtained in the limit of criticality.
Initial measures that are smooth in $\rm U$ with nonzero value outside the
$x=0$ line belong to the basin of attraction of $\mu^G$,
and the initial measures with densities $\phi(y)\delta(x)$ with
$\phi(y)={\cal O}(y^\eta)$ for $y\ll 1$ belong to that of $\mu^G_\eta$.
Note that when deriving Eq.\ (\ref{kfg}) it was assumed that we start with the
Lebesgue measure.
The condition for criticality is again that the escape rate $\kappa^G$
equals to the positive Lyapunov exponent of the saddle point at the origin,
which in case of map (\ref{mbi}) with (\ref{fl1},\ref{fl2})
is $\log(R J^G)$.

Let us turn now to the natural measures.
Still assuming that $\kappa^F$ and $\kappa^G$ are positive
it can be seen that the natural measure corresponding
to $\mu^F_A$ becomes degenerate similarly as in the 1D map,
namely, its density is $\delta(x)\delta(y)$.
This follows from the fact that this natural measure is dominated by the 
trajectories getting close to the saddle point in $x$ direction.
Since they spend most of their lifetime in that vicinity and the saddle is
attracting in $y$ direction with the rate $\kappa^F-\log J^F=\kappa^G$ they
are very close to the saddle in most of their lifetime also in $y$ direction.
%
Numerical approximation of the density of this natural measure,
presented in Fig.\ 4a, also shows this concentration to the fixed point at
the origin,
while the repeller itself does not degenerate in a similar way (see Fig.\ 4b).
Conversely to $\mu^F_A$,
%
(both in $\kappa^G=0$ and $\kappa^G>0$ cases)
the natural measure corresponding to $\mu^F_B$ is distributed
on the whole invariant set except along the line $x=0$.

\begin{figure}
\noindent\hspace*{-1cm}\epsfbox{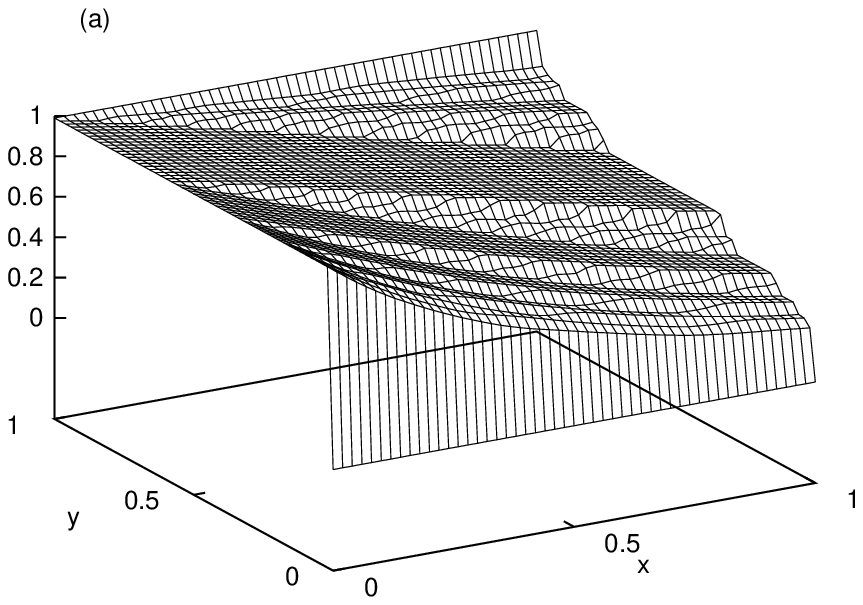}
\end{figure}
\begin{figure}
\hspace*{-1cm}\epsfbox{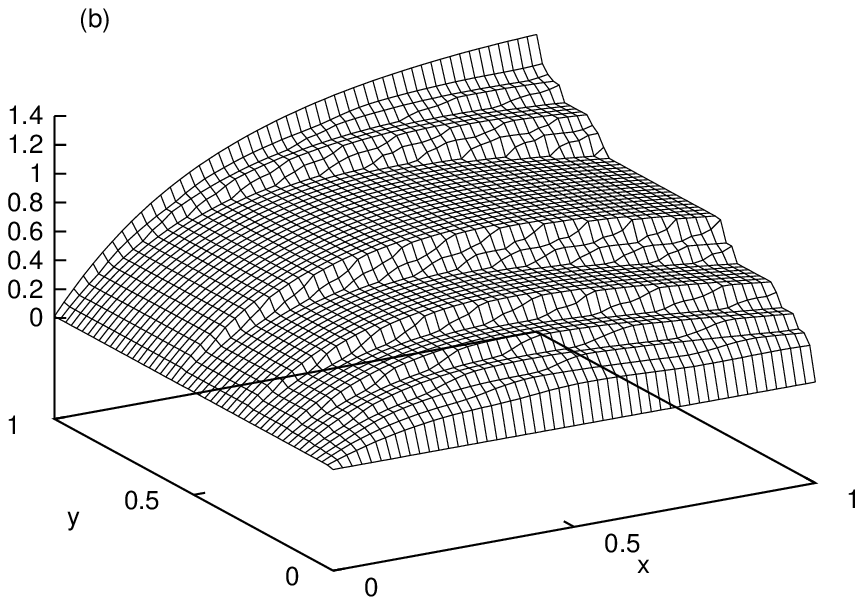}
\end{figure}
\begin{figure}
\hspace*{-1cm}\epsfbox{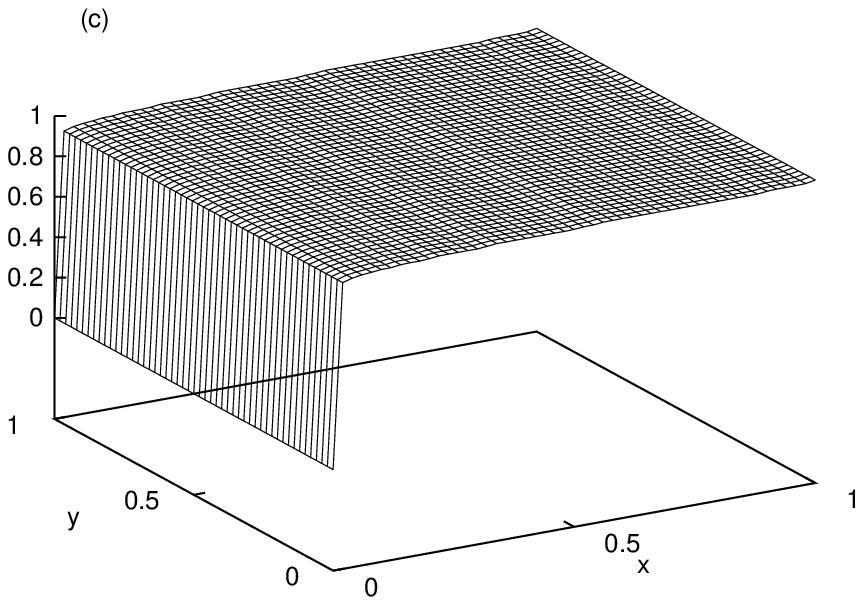}
\caption{The derivatives of
the conditionally invariant measures in the critical state.
The parameters are the same as in Fig.\ 2, except that $d=1$.
(a) $\mu^F_{Ax}(x,y)=\frac{\de}{\de x}\mu^F_A([0,x]\times[0,y])$,
(b) $\mu^F_{Bx}(x,y)=\frac{\de}{\de x}\mu^F_B([0,x]\times[0,y])$,
(c) $\mu^G_y(x,y)=\frac{\de}{\de y}\mu^G([0,x]\times[0,y])$.
The subscripts $A$ and $B$ refer to different initial distributions
as defined in the text.}
\end{figure}

\begin{figure}
\noindent\hspace*{-1cm}\epsfbox{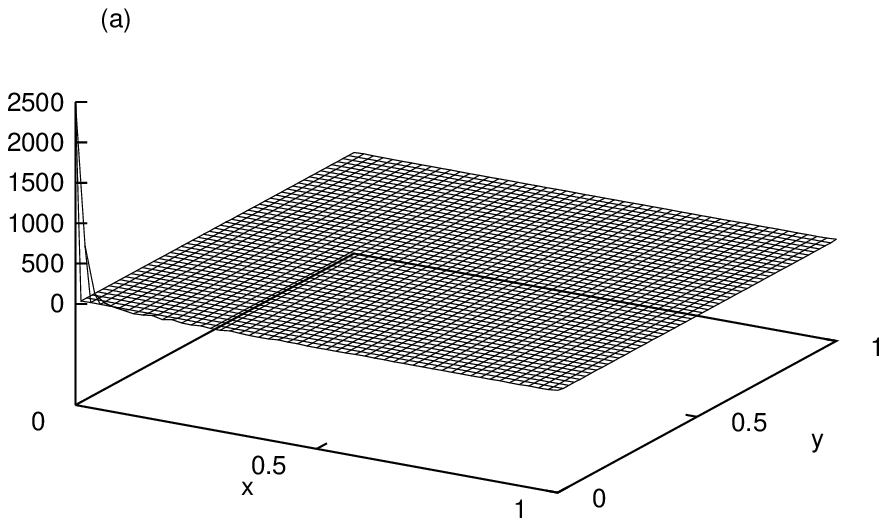}

\noindent\epsfbox{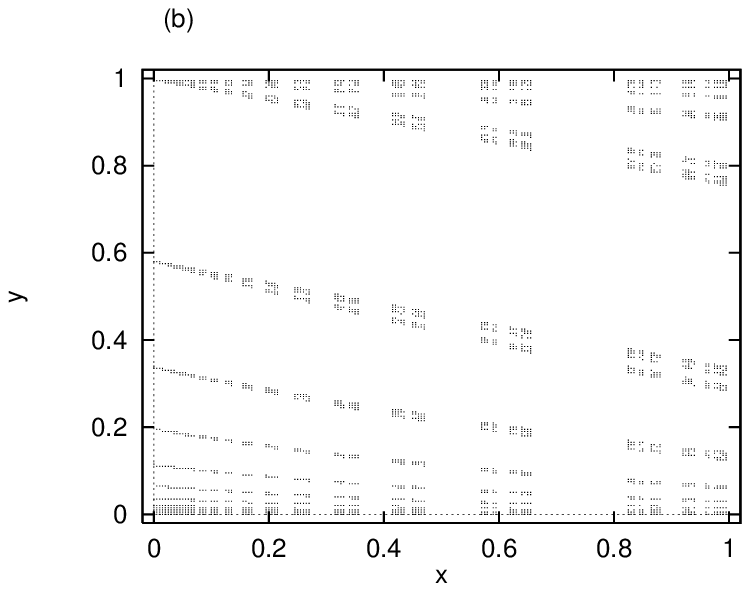}
\caption{Further numerical results in the critical state for the same map
and parameters as in Fig.\ 3a:
(a) the approximate density of the natural measure corresponding to
$\mu^F_A$,
(b) the repeller in finite resolution.}
\end{figure}

Finally we briefly discuss the cases when one of $\kappa^F$ and $\kappa^G$
is zero.
In case of $\kappa^G\rightarrow 0$ the measure $\mu^G$ conditionally invariant
under the iterations by $G$ becomes truly invariant and the related natural
measure coincides with it in the limit.
The behavior of the natural measure can be seen by noticing that in
case $\kappa^G=0$ there is no attraction in $y$ direction at $x=0$.
The fixed point at $x=0, y=0$ and every other point of
the line segment at $x=0$ becomes a marginal fixed point
(they form a fixed line).
Since in the present case $F^n{\rm U}\cap {\rm U}={\rm U}$ and
$F^{-n}{\rm U}\cap {\rm U}$ is uniform in $y$ direction the natural measure
concentrated in $x=0$ is distributed in $y$ direction evenly,
similarly to the conditionally invariant measure of $G$ at $\kappa^G>0$.
Consequently the density of the measure invariant under $G$ is
$\{ \delta(x); 0\le y\le1 \}$.
The dynamics generated by $G$ is strongly intermittent, which is, however,
unusual in the
sense that the measure is not concentrated into a fixed point or a periodic
orbit but to a line segment.
It means that at the intermittent recurrences of the "regular motion" the
values of variable $y$ are evenly distributed between 0 and 1,
while values of $x$ are always near zero then.

In the limit $\kappa^F$ goes to zero one has to follow the development of the
measures of type $A$ and $B$, as well.
One can easily convince oneself, that the measure $\mu^F_A$ remains unaltered
but, of course, becomes invariant.
The related natural measure (entirely concentrated in the fixed point at the
origin) looses its connection to $\mu^F_A$ when $\kappa^F\rightarrow 0$,
i.e.\ the measure concentrated in the fixed point can not be reached by
iterations starting with a smooth measure.
Furthermore $\mu^F_B$ (and the natural measure related to it) becomes
identical to $\mu^F_A$ in this limit.
This measure represents a weakly intermittent dynamics.

\section{Transient Diffusion}

If escape is possible typical long trajectories show a chaotic motion
in a finite duration after which they escape.
The chaotic time period is unlimited
and the behavior of long trajectories is determined by an invariant set.
In the extended system the transiently chaotic motion leads to
transient diffusion.
Such transient diffusion has been studied in one-dimensional maps 
and it has been shown that nonlinearity has strong effects on it\cite{7}.

We turn now to investigate the transient diffusion in the framework of
dynamics generated by $F$ and $G$,
taking into consideration the variable $S$ in
(\ref{mb}) and (\ref{mbi}).
Since the diffusion coefficient is defined in the limit of infinitely long 
trajectories it is an average taking the initial values
according to the natural measure $\nu$.
It is given by\cite{7,26} 
\begin{equation}
D=\lim_{t\rightarrow\infty}
\frac{\langle(S_t-\langle S_t\rangle_\nu)^2\rangle_\nu}{2t}\;,\label{d}
\end{equation}
where  $\langle S_t\rangle_\nu$ is the average drift and we start with $S_0=0$.

First consider the noncritical situation, when $\nu$ is obtained
starting with the Lebesgue measure.
Choosing first the map $F$ and starting with a point $(x_0,y_0)$ from the cell
$S_0=0$ we obtain
\be
(x_t,y_t,S_t)=F^t(x_0,y_0,0)\;.
\ee
Since $\nu$ is invariant distributing initial points $(x_0,y_0)$
according to it one observes the same distribution for $(x_t,y_t)$.
Obviously
\be
G^t(x_t,y_t,S_t)=(x_0,y_0,0)\;,
\ee
and due to translational invariance
\be
G^t(x_t,y_t,0)=(x_0,y_0,-S_t)\;.
\ee
Furthermore recalling the result in Section IV, that $\nu$ is the same for
$F$ and $G$, one concludes that the definition (\ref{d}) gives
the same diffusion coefficient for the maps $F$ and $G$.

Let us turn now to the critical state.
We have seen that in case of map $F$ we have to consider two natural measures.
The first one is related to the Lebesgue measure
and is entirely concentrated into the fixed point.
Consequently, the diffusion coefficient is zero.
The other one leads, however, to a nonzero diffusion coefficient.

Regarding the map $G$ the natural measure related to $\mu^G$
leads also to $D=0$.
That means changing a control parameter the diffusion coefficient
tends to zero when approaching criticality.
For the map $F$ a jump can occur in $D$ when reaching criticality,
while for $G$ this is not the case.

\section{Summary and Discussion}

In this paper we have followed the development of conditionally
invariant and related natural measures when altering a control parameter
of the generalized multibaker map. 
It has been investigated which conditionally invariant measure
we arrive at if the Lebesgue measure is the initial one.
It can be easily seen that away from the critical state (i.e.\ $v'(0)<1$)
the same conditionally invariant measure is obtained by starting the iteration
with such a measure that decreases when approaching the line segments
$x=0,1,\; 0\leq y\leq 1$
at least as fast as the Lebesgue measure (the condition can be
weakened, so the complete basin of attraction is larger).
We call a conditionally invariant measure typical if its basin of attraction is
specified by limitation of its scaling only from above
at both of the above mentioned line
segments, thereby its basin of attraction is large with respect to its
exponents at $x=0$ and $x=1$.
In particular a typical conditionally invariant measure has
such smooth distributions in its basin of attraction 
that are concentrated in the interior of the unit square.
In this sense the conditionally invariant measure developed from the Lebesgue
measure is typical away from the critical state.
At the critical state, however, this is no more the case if we treat the
map $F$.
Starting with
the Lebesgue measure one arrives at $\mu_A^F$, but its basin
of attraction is restricted in the way that the initial measure
should start linearly at $x=0$ ($0\leq y\leq1$) if it behaves at $x=1$
as $\mu^0(1)-b(1-x)^\tau$, $\tau>1/\omega$ 
or it can have more general scaling possibilities at $x=0$ if however
its asymptotics at $x=1$ ($0\leq y\leq 1$) is fixed. So according to
our definition above $\mu_A^F$ is not typical.
One should point out
that there are continuously many non-typical conditionally
invariant measures at and away from the critical state, as well,
whose discussion,
however, is beyond the scope of the present paper.
In case of 1D maps 
these measures have been investigated in detail in Ref.\ 25. 
Their appearance is connected to the existence of singular
eigenfunctions of the Frobenius-Perron operator first pointed
out and studied by MacKernan and Nicolis in case of piecewise
linear maps\cite{27}. 
The measure $\mu_A^F$ is 
unique among the other non-typical measures at the
critical state because it is the limit of typical ones
when approaching the critical state by changing the control
parameter.
Furthermore this conditionally invariant measure is obtained
when one starts with the Lebesgue measure.
At the critical situation $\mu_B^F$ is typical, its
basin of attraction contains distributions specified in Sec IV\@.
Summarizing, if one changes the control parameter the following
behavior emerges in case of the map $F$
starting with a distribution with density concentrated
in the interior of the unit square:
away from the critical state the measure approaches the same
conditionally invariant measure as the Lebesgue measure while arriving
at the critical state the measure $\mu_B^F$ is obtained, which does not
have the Lebesgue measure in its basin of attraction.
This leads to a jump in the diffusion coefficient when
reaching the critical state.

Concerning the map $G$, away from the critical state the same
properties apply as for $F$.
The typical conditionally invariant measure remains typical,
however, in the limit of the critical state with the basin
of attraction containing distributions which are not concentrated
on the $x=0,\; 0\leq y\leq 1$ line segment.
Among the non-typical conditionally invariant measures
existing for the map $G$
we mentioned in Section IV those which are concentrated on 
the line segment mentioned above.
The behavior of the typical conditionally invariant measure
has led to a vanishing diffusion coefficient when approaching
the critical state in case of the map $G$.

There is an aspect of our investigation which we would like to emphasize.
It is generally assumed in case of transient chaos that the initial measure is 
the Lebesgue one.
It is then a basic question, what is the basin of attraction of the 
conditionally invariant measure which is arrived at when one starts with the 
Lebesgue measure.
Our results show that this conditionally invariant measure has a considerable 
basin of attraction in general, an exception is provided by the critical state
in case of map $F$.
Finally one has to point out that other initial measures outside the basin of 
attraction of this conditionally invariant measure might be realized not only in 
numerical, but also in laboratory experiments.

Taking the limit of permanent chaos along the "thermodynamic path"
maintaining the critical state
we have discussed the intermittent behavior of the dynamics and
pointed out also unusual properties.
The intermittent
dynamics can be contrasted to that generated by the map
$F$ with the $\mu_A^F$ measure in the critical state. 
As shown in Section IV the related invariant (natural)
measure is entirely concentrated in the origin, a feature
shared also by strong intermittency in its usual form, 
which means that in both
cases the trajectories spend the majority of time near the
origin.
The transient chaos corresponding to the $\mu_A^F$
measure is qualitatively different, however, from intermittency
(at least not very close to the permanent chaos limit)
since the majority of trajectories go away from the vicinity of the
origin without exhibiting regular-like sequences characteristic
to intermittency and they escape.

\acknowledgements

This work has been supported
by the Hungarian National Scientific Research Foundation under Grant 
Nos.\ OTKA T017493 and OTKA F17166. 
Authors thank H. Lustfeld for helpful discussions.

\end{document}